\documentclass[numberedappendix]{emulateapj}
\usepackage{natbib}
\shorttitle{Galactic Wind Signatures}
\shortauthors{Kawata and Rauch}

\begin{document}


\title{Galactic Wind Signatures around High Redshift Galaxies}

\author{Daisuke Kawata\altaffilmark{1,2}
and Michael Rauch\altaffilmark{1}, 
}

\altaffiltext{1}{The Observatories of the Carnegie Institution of Washington,
 813 Santa Barbara Street, Pasadena, CA 91101
\email{dkawata, mr@ociw.edu}}
\altaffiltext{2}{
 Swinburne University of Technology, Hawthorn VIC 3122, Australia
}

\begin{abstract}
We carry out cosmological chemodynamical simulations with different
strengths of supernova (SN) feedback and study how galactic winds 
from star-forming galaxies affect the features of hydrogen (HI) 
and metal (CIV and OVI) absorption systems in the intergalactic medium
at high redshift.
We find that the outflows tend to escape to low density regions,
and hardly affect the dense filaments visible in HI absorption.
As a result, the strength of HI absorption near galaxies 
is not reduced by galactic winds, but even 
slightly increases. We also find that a lack of HI absorption 
for lines of sight (LOS) close to galaxies, as found by Adelberger et~al., 
can be created by hot gas around the galaxies induced by 
accretion shock heating. In contrast to HI, metal absorption systems are 
sensitive to the presence of winds. The models without 
feedback can produce the strong CIV and OVI absorption lines in LOS
within 50 kpc from galaxies, while strong SN feedback is capable of
creating strong CIV and OVI lines out to about twice that distance.
We also analyze the mean transmissivity of HI, CIV, and 
OVI within 1 h$^{-1}$ Mpc from star-forming galaxies. The probability 
distribution of the transmissivity of HI is independent of the 
strength of SN feedback, but strong feedback produces LOS with 
lower transmissivity of metal lines. 
Additionally, strong feedback can produce strong OVI lines even 
in cases where HI absorption is weak. We conclude that OVI is probably 
the best tracer for galactic winds at high redshift.
\end{abstract}


\section{Introduction}
\label{sec-intro}

Supernova (SN) explosions are thought 
to be capable of ejecting part of the interstellar medium (ISM) from
galaxies. Such outflows
are often called ``galactic winds''
\citep{ja71,mb71,vcb05}. Galactic winds are believed to be
an important mechanism for enriching the intergalactic medium (IGM)
\citep{si77,ahs01,mfr01,sfm02,cno05} and are thought to play a crucial role 
in shaping the mass-metallicity relation
of galaxies \citep[e.g.][]{rl74b,ds86,ay87,bg97,kg03a}, and in heating
the IGM \citep{io86}.
Galactic winds have been observed in
local star-forming galaxies \citep[e.g.][]{ls63,cm98,oti02},
where their outflow morphology and kinematics have been
extensively studied \citep[e.g.][]{cm05,rvs05}.
Some local galaxies show outflows to about 20 kpc
\citep{vsrbc03}.
Such outflows are expected to be more common at high redshift,
where star formation is more active ($z>1$) 
\citep{mfd96}.
Galactic winds are also believed to terminate star formation
in ellipticals \citep{mb71,kg03a} and have been invoked 
at high redshift to explain
the high age of their stellar populations \citep{ka97,lhf05,kvf06}.
Theoretical studies suggest that  for progenitors of
disk galaxies the gas outflow due to SN heating at high redshift
can effectively suppress star formation, leading to a
less dense stellar halo \citep{bkgf04a} and 
a larger disk \citep{sgp03,rysh04,gwm06} at $z=0$.
Therefore, observing galactic wind signatures at high redshift 
may elucidate an important ingredient in the formation of galaxies.

The observational studies at high redshift have uncovered 
evidence that outflows from star-forming galaxies at high redshift
may be common, as has been shown for Lyman break selected galaxies
\citep[e.g.][]{pks98,otk03,sspa03}. 
These results have contributed to the debate about how the outflows
from star-forming galaxies may affect the intergalactic medium (IGM).
\citet{rsb01} uncovered evidence for
repeated injection of kinetic energy into higher density, CIV absorbing gas,
possibly driven by
recent galactic winds.  In a study of the {\em lower density}, general
Lyman alpha forest, \citet{rsbc01} found that most that of the HI
absorption systems lack signs for being
disturbed by winds, and derived upper limits on the filling factor of
wind bubbles. \citet{ssr02}
surveyed the properties of strong OVI absorption systems at high
redshift and proposed that
the apparent temperatures and the kinematics of the OVI gas as well as
their rate of incidence
could be explained if massive Lyman break galaxies are driving winds
out to 50 proper kpc.

\citet{assp03,ass05} studied the absorption line features in the spectra
of background QSOs whose line of sight (LOS) passes close to
Lyman-break selected
star-forming galaxies. They found a deficit of neutral hydrogen near
these galaxies out to 0.5 $h^{-1}$ comoving Mpc, accompanied by a
surplus of HI beyond that radius, and suggested that 
most Lyman break galaxies may reside bubbles where superwinds have
depleted the HI in the interior and 
piled up more neutral gas beyond the hot bubble. The more recent, more
statistically
significant of these studies \citet{ass05}
however, does not support this claim, but still appears to show  a significant
fraction (about 7 out of 24) of the LOS exhibiting weak
or no absorption within 1 $h^{-1}$ comoving Mpc. Numerical
simulations of the IGM 
without SN feedback predict a much lower fraction of such weak
absorption systems near the galaxies \citep{kwdk03}, a fact that may
conceivably be explained if at least some galaxies have outflows
destroying HI in the IGM in their vicinity. 
\citet{assp03,ass05}  also show that
there is a correlation between the spatial distribution
of CIV absorption lines and the star-forming galaxies, with
the strongest CIV absorption lines being observed at the LOS 
closest to the galaxies ($\sim80$ proper kpc). This result may indicate
that outflows related to recent star formation activity have enriched the IGM
locally.
Other evidence for the association of CIV with galaxies has been reported
\citep{psa06,ssrb06}. \citet{spa06}
compared the observed LOS correlation functions of CIV and SiIV with
analytic outflow models and concluded that
the observed correlation can be formally explained if there are 
outflows with a scale of about 2 comoving Mpc from large
galaxies whose stellar mass is about $10^{12}$ M$_{\sun}$.
However, the large range of the individual wind bubbles required to explain
the CIV- galaxy correlation is puzzling. Theoretical arguments suggest
that the clustering of CIV with Lyman break
galaxies is not necessarily proof that those same galaxies produced the metal
enrichment \citep{pm05,es05}, and
the relation between metals in the IGM and galaxies clearly needs
further study.

 Cosmological numerical simulations have proven a useful tool for 
understanding metal absorption lines \citep{rhs97}.
Comparisons between the observations and the statistics of metal
absorption lines derived from numerical simulations have mostly been concerned
with measurements of the metallicity and ionization state
of the IGM \citep[e.g.][]{dhhkw98,ast02,sak03,ask04}. The relation 
between star-forming galaxies and absorption line features
have recently been studied by a number of authors \citep{chsww02,kwdk03,
bfm03,kmco06,tb06}, with several papers \citep[e.g.][]{chsww02,kmco06,tvk02}
considering observable effects of galactic outflows on the Lyman alpha
forest absorption lines.
These studies generally have concluded that such outflows hardly affect
the strength of the HI absorption lines, because the winds tend to escape into
less dense regions and do not impact the IGM where the density is
high enough to produce HI absorption lines.

So far, most studies based on full-blown cosmological
numerical simulations have focused on HI absorption lines. Although
there have been a number of papers discussing the origin and
properties of metal absorption lines
\citep[e.g.][]{rhs97,tvk02,ash05,tb06,od06}, the relation between metal
absorption line systems and outflows from the coeval galaxy population
has remained largely unclear, in particular as the effect of winds on
the physical properties
of metal lines it is not well known. Nevertheless, the fact that the
metallicity is expected to be 
increased by galactic outflows, and the availability of multiple
transitions and several heavy elements
with a different enrichment history should make metal lines a
potentially much more useful tracer
of galactic winds than neutral hydrogen.

 The current paper studies the properties of both HI and metal absorption
lines, and looks for observational signatures of galactic winds.
To this end, we run cosmological simulations with
the original version of the Galactic Chemodynamics Code, {\tt GCD+}
\citep{kg03a} which is capable of tracing the chemical evolution of the IGM
and galaxies self-consistently. We carry out simulations
with different strengths of SN feedback, and compare the features
of the HI and metal absorption lines between the simulations,
attempting to identify features sensitive
to the presence of galactic winds. 

The following section will explain our method including the 
description of the numerical simulations with {\tt GCD+}
and analysis of absorption lines. Unlike previous studies,
we follow different heavy elements separately, and
take into account the abundance evolution for the different elements
when creating fake QSO spectra from the simulation.
This is important because the different elements come 
from different types of SNe or are due to mass loss from inter-mediate mass stars
with different life-times. Section \ref{sec-res} shows
our results. First, Section \ref{sec-afag} focuses on 
one galaxy in the simulation volume, and compares the 
absorption line features around the galaxy among models
with different SN feedback strengths. Then, in Section \ref{sec-mtrans}
we discuss the results more quantitatively using artificial QSO spectra
in 1000 random LOS. Section \ref{sec-conc} summarizes our conclusions.

\begin{figure}
\epsscale{.70}
\plotone{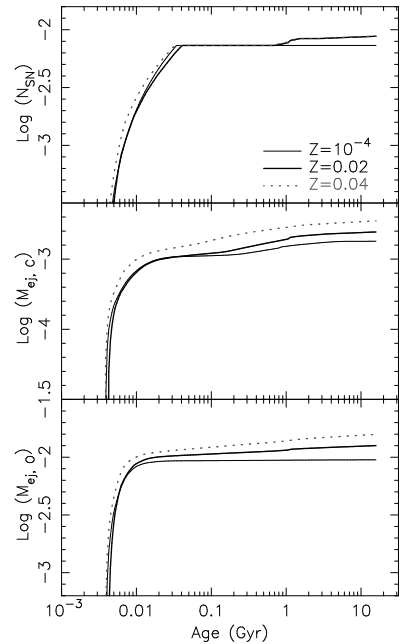}
\caption{
 Number of SN and chemical yields as a function of the age of a star 
particle with a mass of 1 M$_{\sun}$. 
The upper panel shows the total number of SN.
The middle and lower panels present the total ejected carbon and oxygen
masses, respectively. 
The thin solid, thick solid, and gray thick dotted lines indicate the history
of a star particle with the metallicity of 
$Z=10^{-4}$, 0.02, and 0.04, respectively.
\label{fig-tnco}}
\end{figure}

\begin{figure*}
\plotone{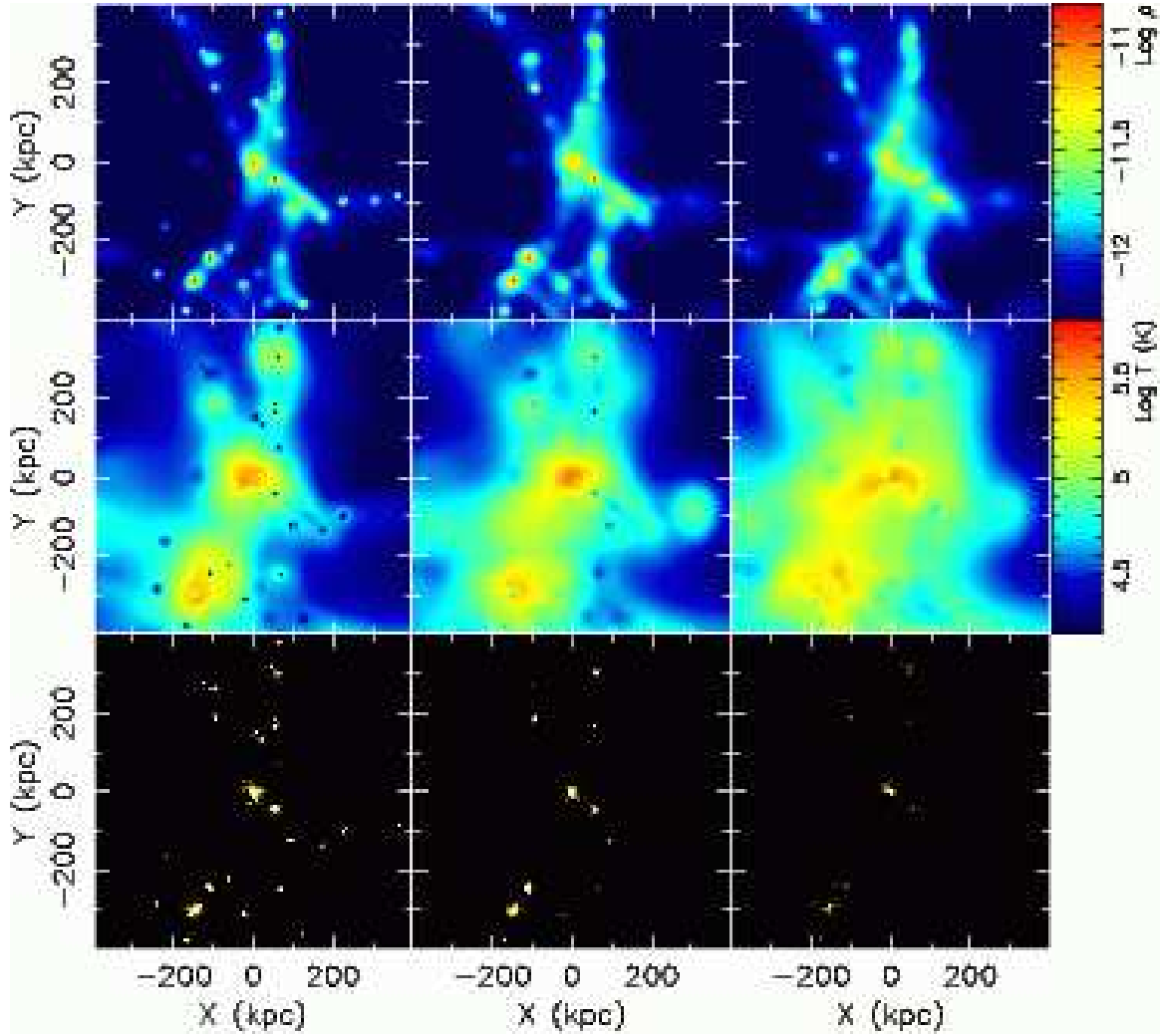}
\caption{
Projected gas surface density (upper) and temperature (middle) map
and star particle distribution (lower)
for models NF (left), SF (middle), and ESF (right).
\label{fig-pmap}}
\end{figure*}

\begin{figure*}
\plotone{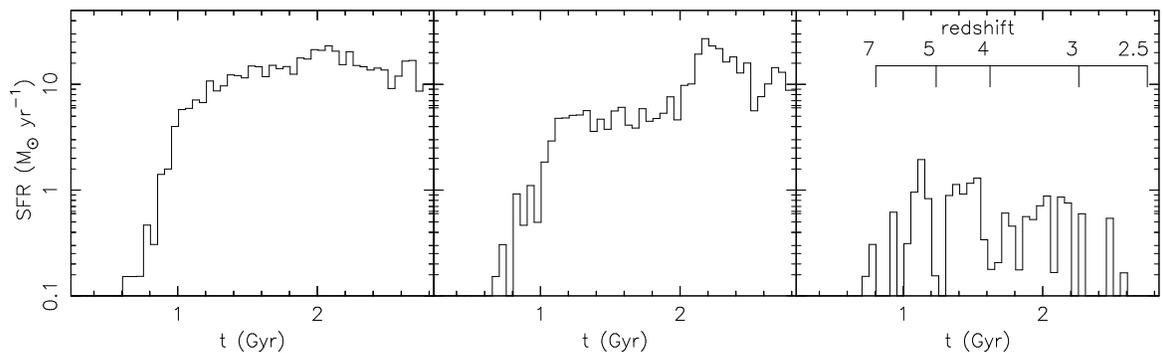}
\caption{
 The history of the star formation rate down to $z=2.43$ for models
NF (left), SF (middle), and ESF (right).
The time of first star formation is different among the three models,
although there should not be any difference before feedback from
stars happens. This difference is because different models are carried out
on different computers, and the star formation model in {\tt GCD+}
uses the random number generator \citep[see][for details]{kg03a} whose 
sequences are different for different simulations.
\label{fig-sfr}}
\end{figure*}

\begin{figure*}
\plotone{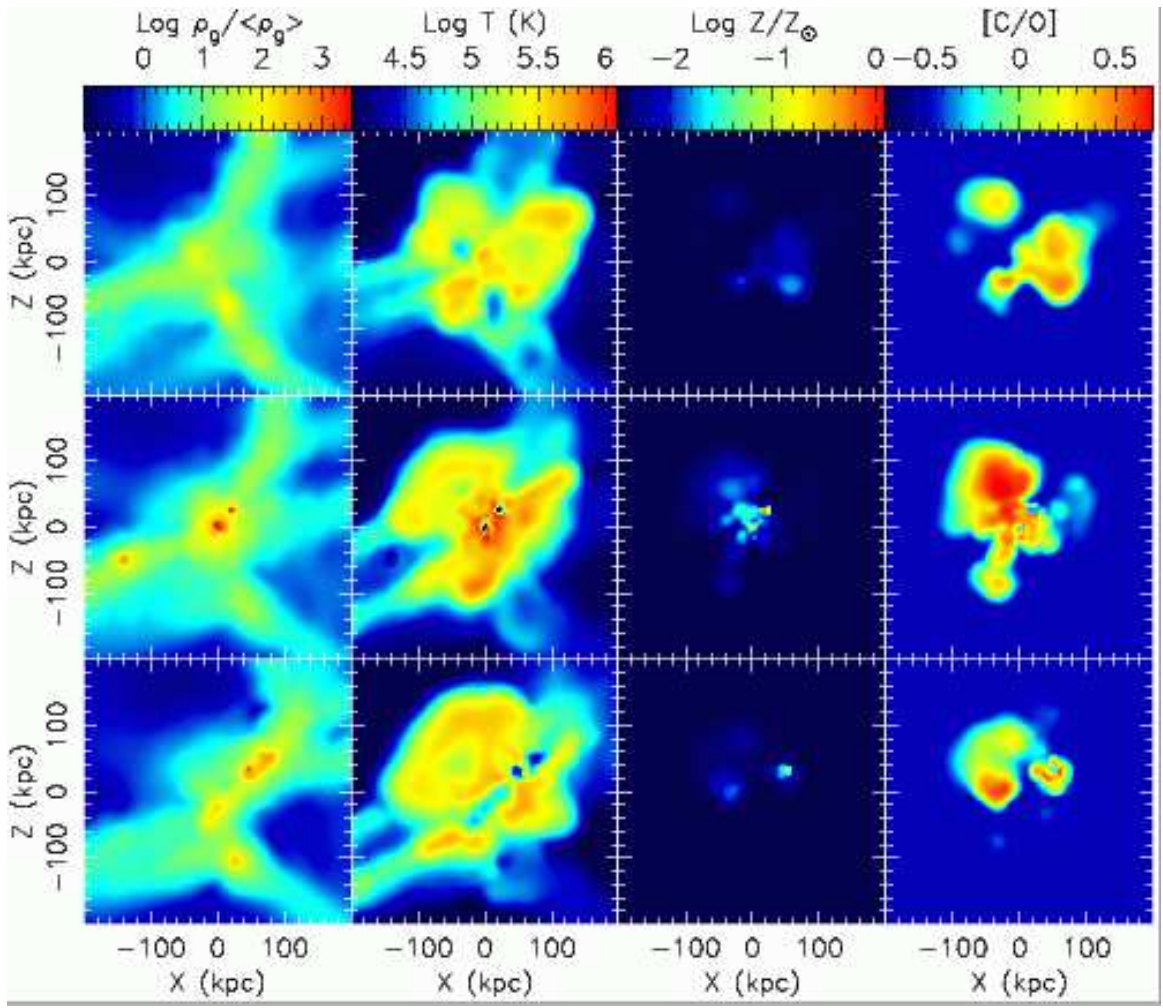}
\caption{
Overdensity, temperature, metallicity, and
[C/O] map (from left to right) for model NF
at Y$=+50$ (upper),
0 (middle) $-50$ (lower) proper kpc, where the biggest galaxy is
set to be at the center.
\label{fig-slmsne0}}
\end{figure*}

\begin{figure*}
\plotone{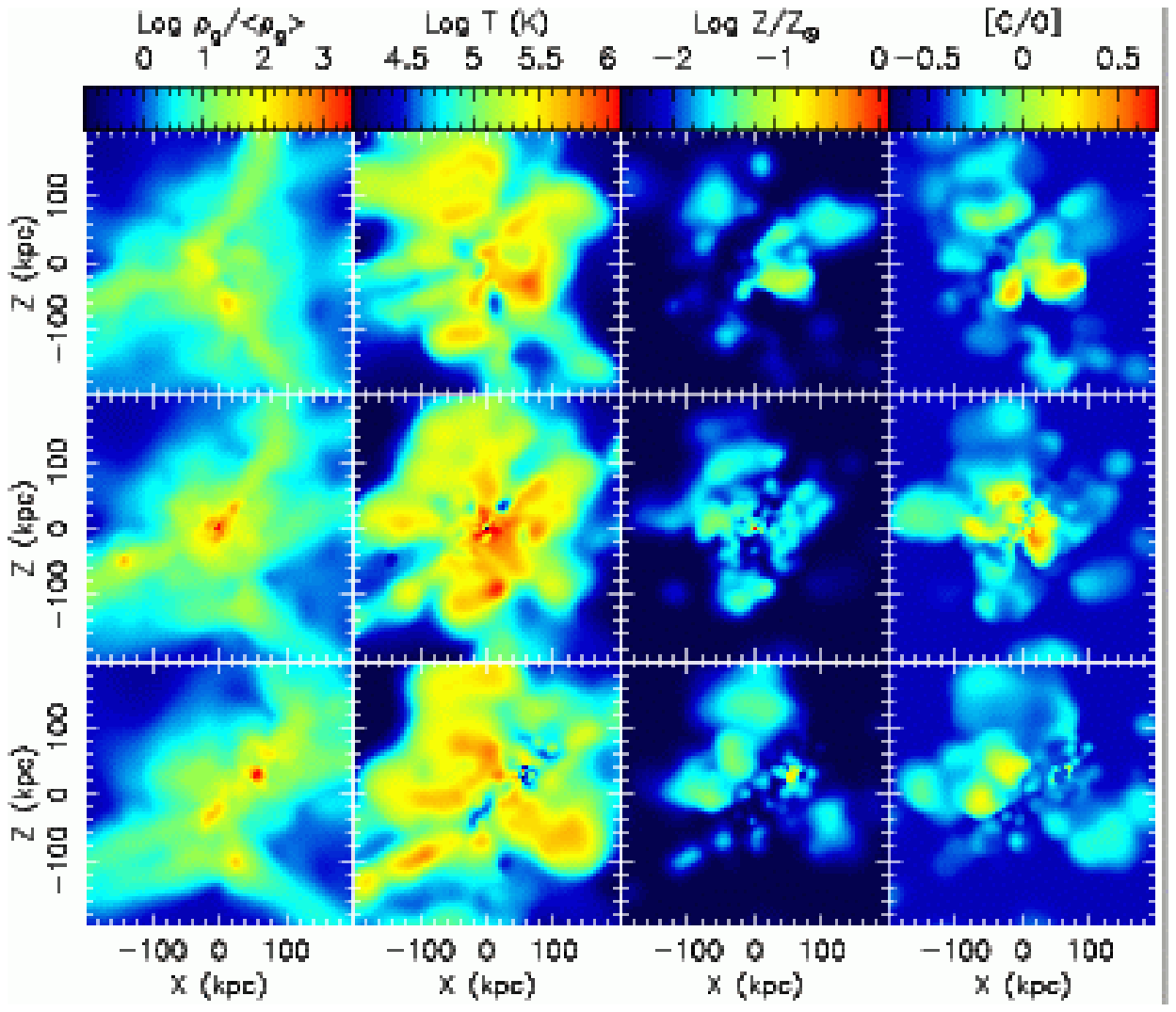}
\caption{
Same as Fig. \ref{fig-slmsne0}, but for model SF.
\label{fig-slmsne3}}
\end{figure*}

\begin{figure*}
\plotone{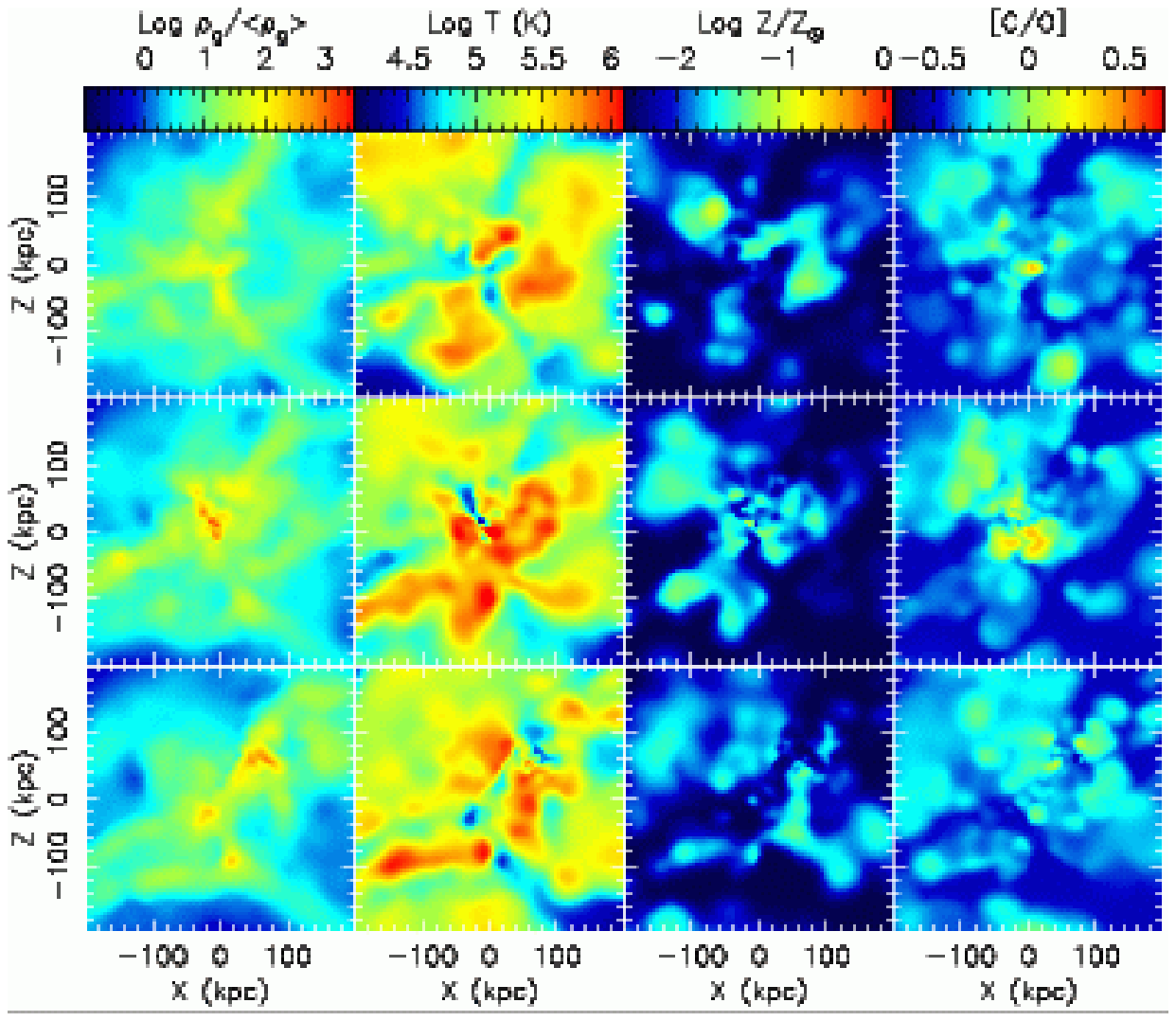}
\caption{
Same as Fig. \ref{fig-slmsne0}, but for model ESF.
\label{fig-slmsne5}}
\end{figure*}

\begin{figure*}
\plotone{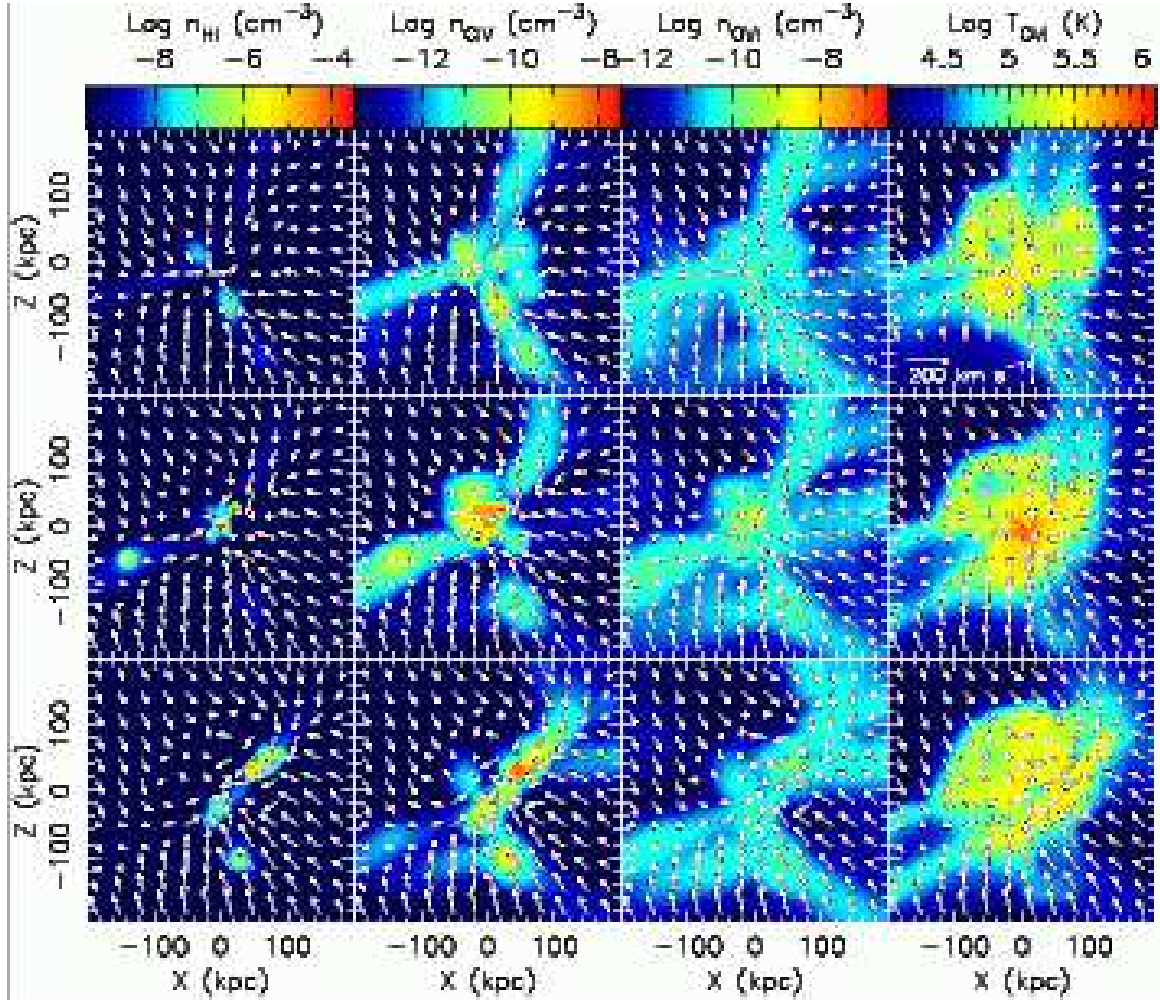}
\caption{
Density of HI, CIV, and OVI and
OVI weighted temperature map (from left to right) for model NF
at Y$=+50$ (upper),
0 (middle) $-50$ (lower) proper kpc, where the biggest galaxy is
set to be at the center. The arrows represent the velocity field
weighted by HI, CIV, OVI, and OVI in the panels from left to right, 
respectively.
The size of arrow corresponds to the amount of velocity,
as indicated in the upper right panel.
\label{fig-slzmsne0}}
\end{figure*}

\begin{figure*}
\plotone{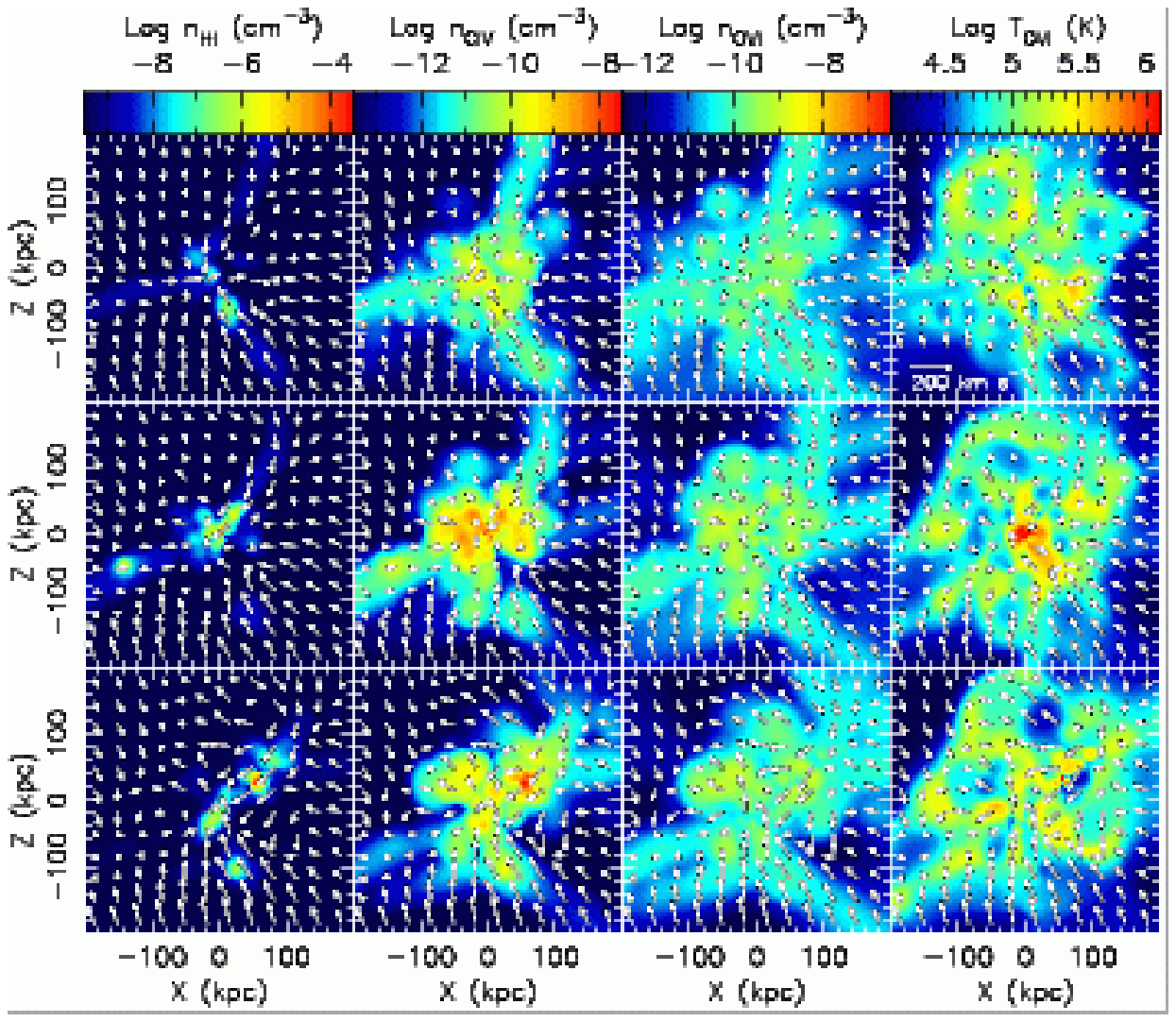}
\caption{
Same as Fig. \ref{fig-slzmsne0}, but for model SF.
\label{fig-slzmsne3}}
\end{figure*}

\begin{figure*}
\plotone{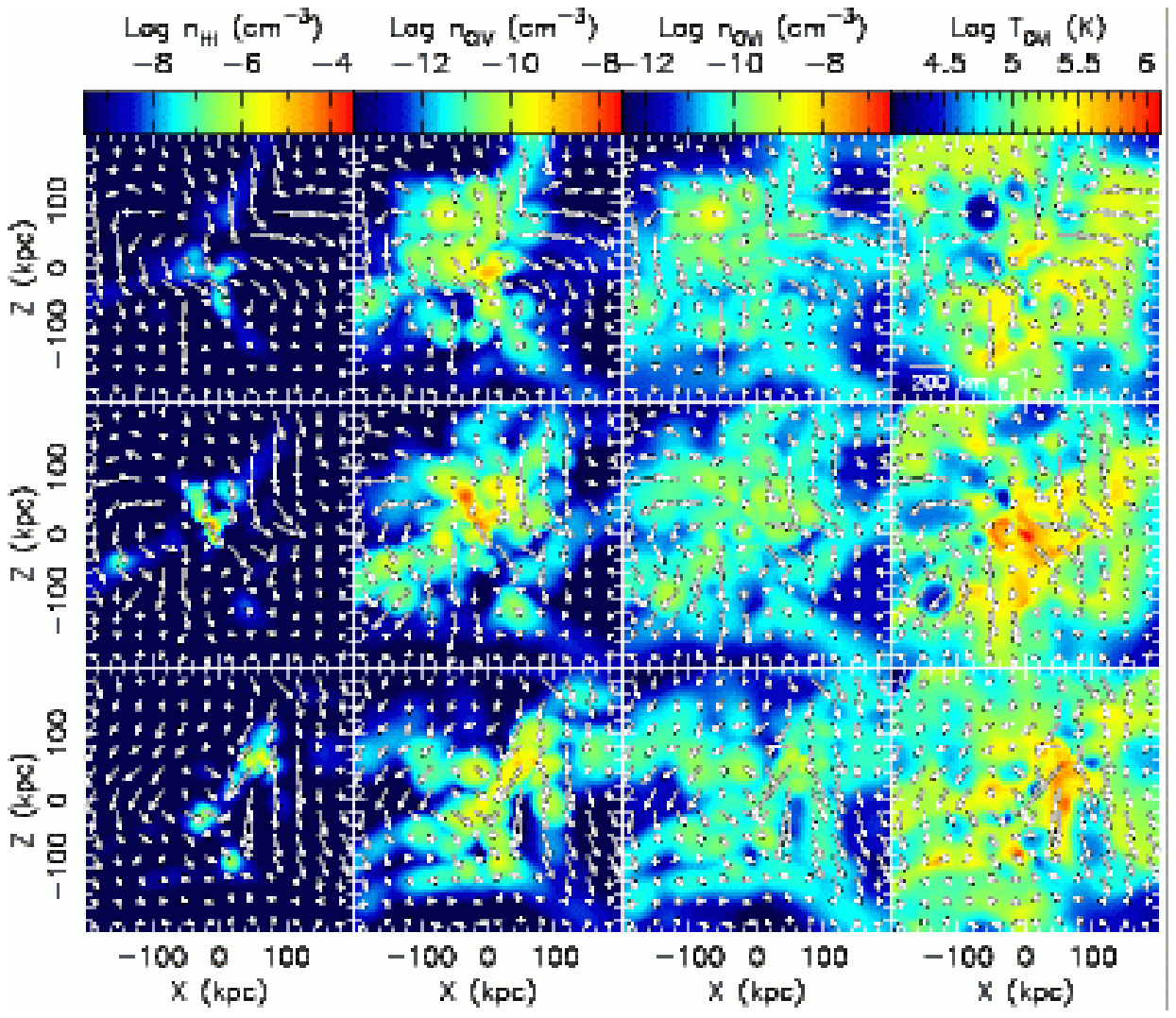}
\caption{
Same as Fig. \ref{fig-slzmsne0}, but for model ESF.
\label{fig-slzmsne5}}
\end{figure*}

\begin{figure}
\epsscale{.80}
\plotone{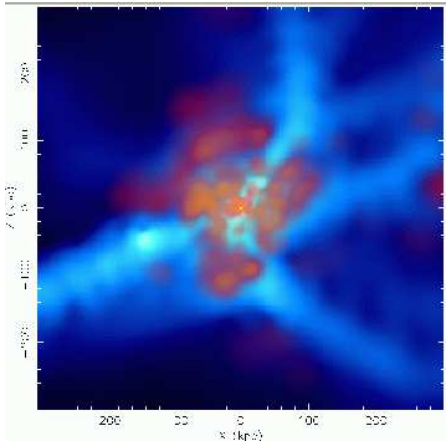}
\caption{
 Composite image of overdensity (blue contour) and metallicity
(red contour) distribution at Y$=0$ for model SF. 
The edges of contours for overdenisty
and metallicity correspond to $\log \rho_{\rm g}/<rho_{\rm g}>=-1$
and $\log Z/Z_{\sun}=-2.5$, respectively.
\label{fig-dZsne3}}
\end{figure}


\begin{figure*}
\plotone{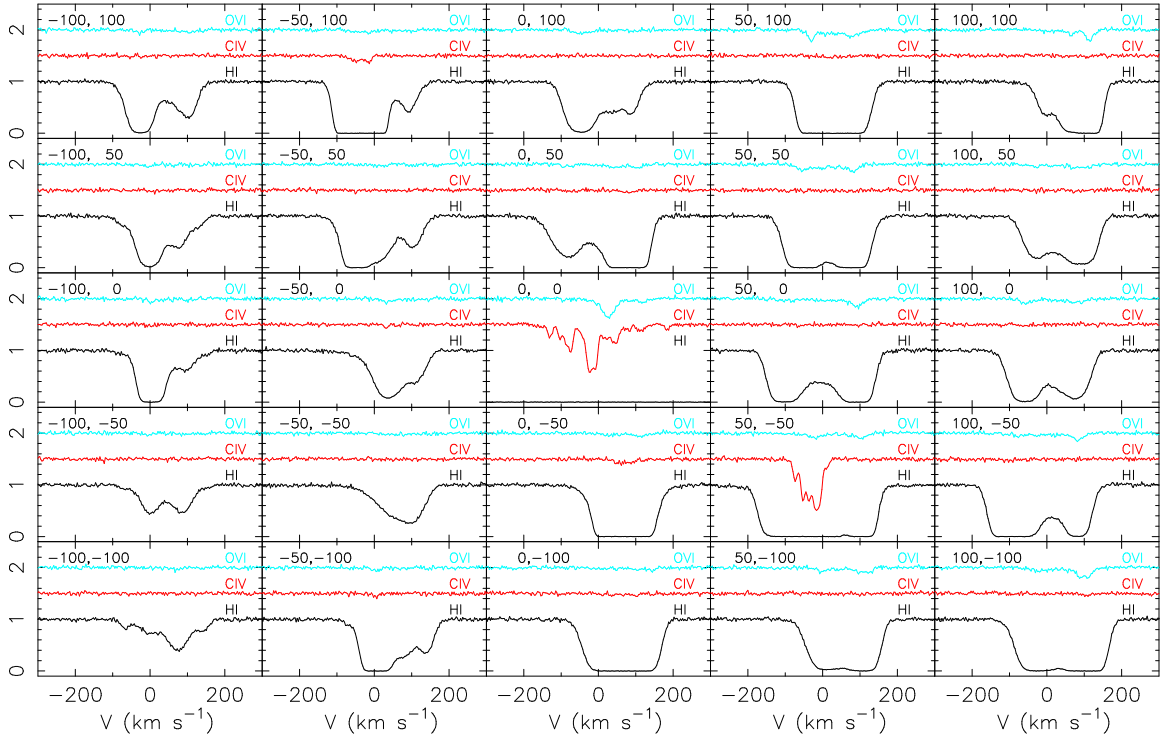}
\caption{
Mosaic of $5\times5$ lines of sight spectra of HI, CIV, and OVI lines
around a galaxy at $z=2.43$ for model NF. 
The central panel corresponds to the position of the galaxy, and
each line of sight is separated by 50 kpc in physical scale. 
The numbers at the upper left corner of each panel present
the (x,y) coordinate (in kpc) for each LOS. 
In the case of 
CIV and OVI only the stronger line of each doublet is shown.
The LOS velocity is adjusted so that the LOS velocity of the galaxy
equals zero.
\label{fig-spgridsne0}}
\end{figure*}

\begin{figure*}
\plotone{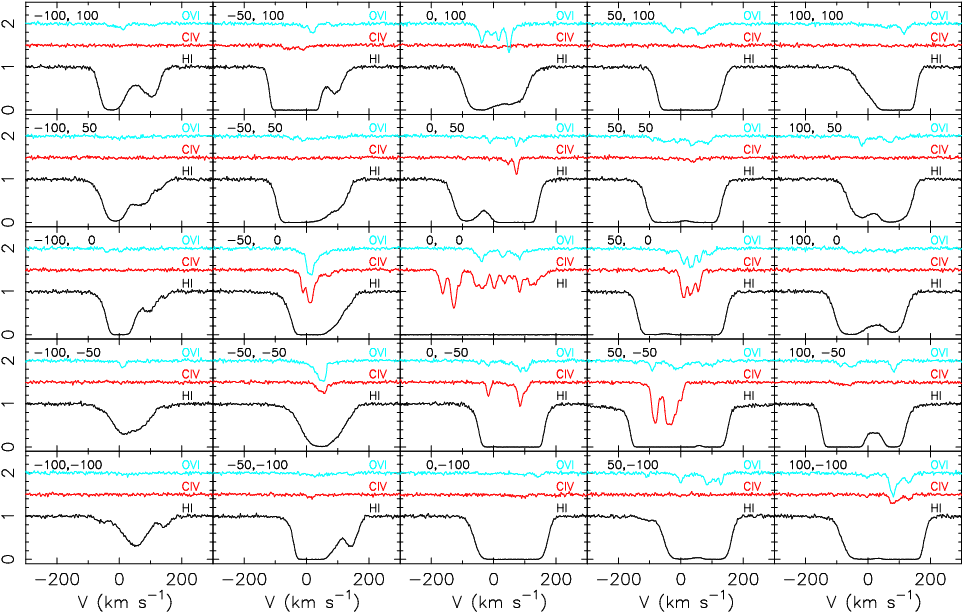}
\caption{
Same as Fig. \ref{fig-spgridsne0}, but for model SF.
\label{fig-spgridsne3}}
\end{figure*}

\begin{figure*}
\plotone{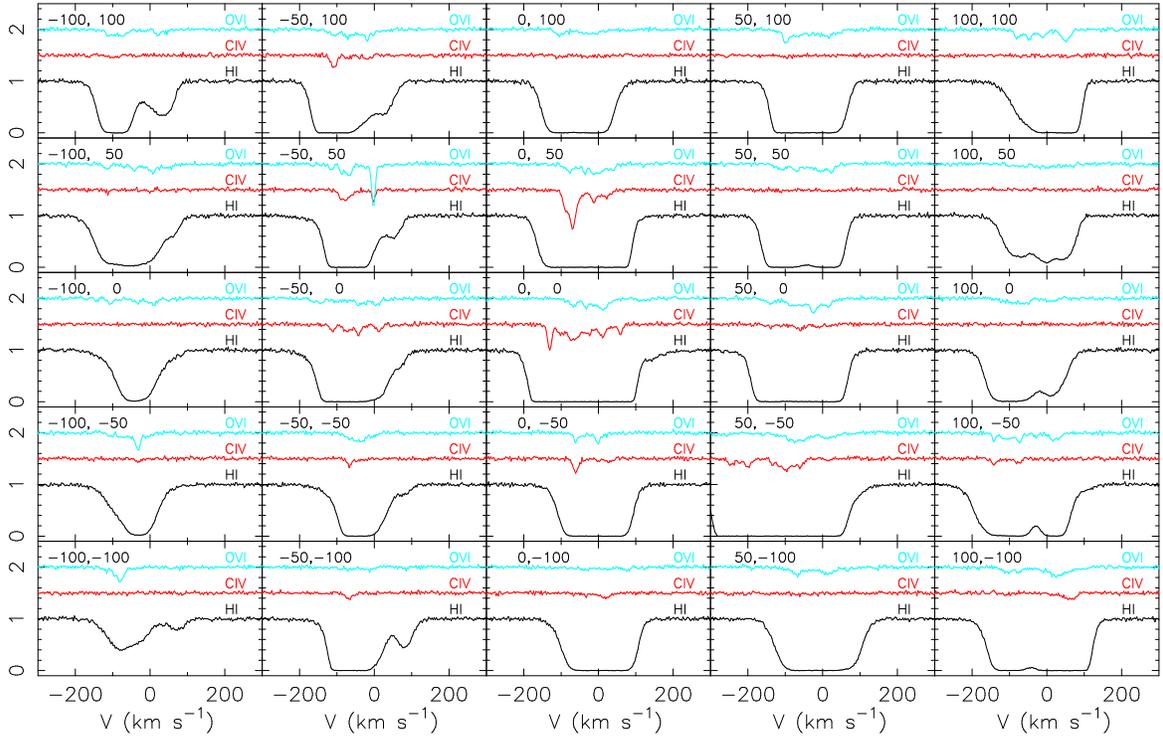}
\caption{
Same as Fig. \ref{fig-spgridsne0}, but for model ESF.
\label{fig-spgridsne5}}
\end{figure*}


\begin{figure*}
\plotone{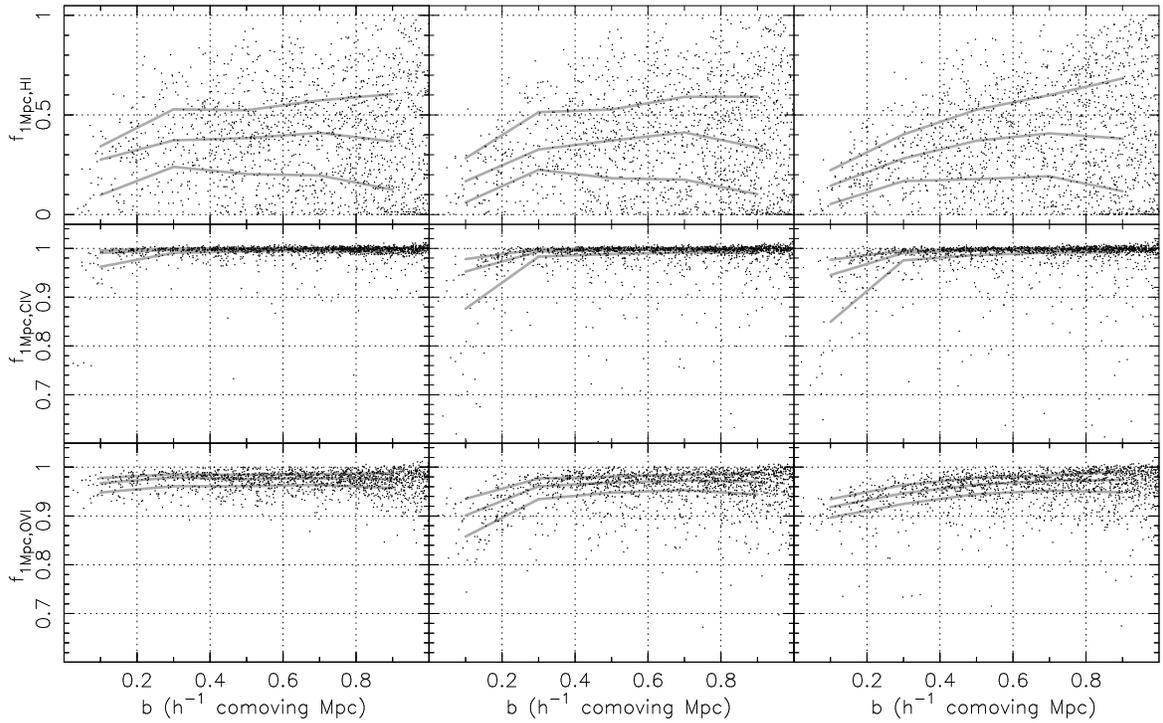}
\caption{
 The mean transmissivity of HI (upper), CIV (middle), and OVI (lower)
for the pixels within 1 $h^{-1}$ comoving Mpc of the galaxies 
as a function of the impact parameter for models NF (left), SF (middle),
and ESF (right). The gray lines indicate median and 25th and 75th 
percentile.
Note that the scale of the y-axis, i.e., $f_{\rm 1Mpc}$ 
is different in each panel.
\label{fig-bfmpc}}
\end{figure*}

\begin{figure*}
\plotone{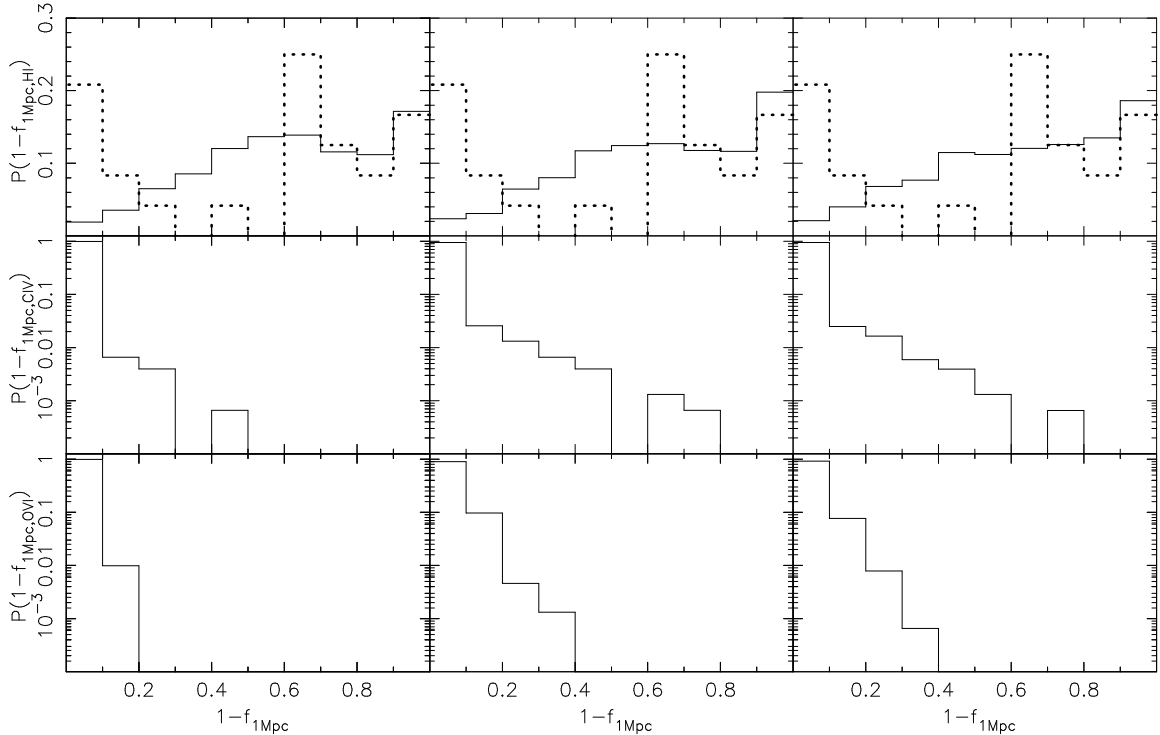}
\caption{
 The probability of the flux decrement $1-f_{\rm 1Mpc}$
of HI (upper), CIV (middle), and OVI (lower), where
$f_{\rm 1Mpc}$ is the mean transmissivity  for the pixels within
 1 $h^{-1}$ comoving Mpc of the galaxies.
 The left/middle/right panel shows the results of model NF/SF/ESF.
The dotted histogram in the upper panels present
the observational results of \citet{ass05}.
\label{fig-fhist}}
\end{figure*}

\begin{figure*}
\plotone{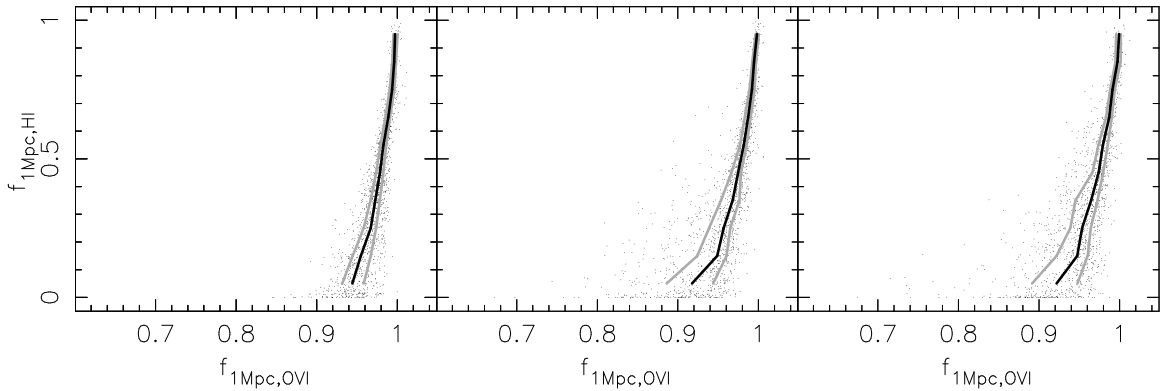}
\caption{
 The mean transmissivity of the galaxies of HI, $f_{\rm 1Mpc,HI}$ 
for the pixels within 1 $h^{-1}$ comoving Mpc 
as a function of the mean transmissivity of OVI, $f_{\rm 1Mpc,OVI}$,
for models NF (left), SF (middle), and ESF (right).
The lines indicate
median (black line) and 25th and 75th percentile (grey lines).
\label{fig-fovi}}
\end{figure*}

\begin{figure*}
\plotone{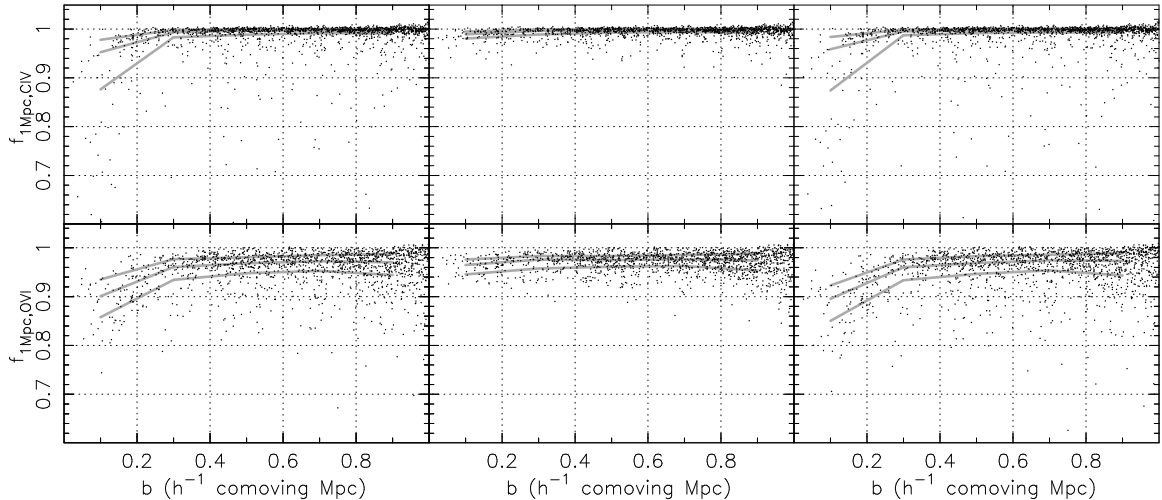}
\caption{
 The mean transmissivity of CIV (uppera), and OVI (lower)
for the pixels within 1 $h^{-1}$ comoving Mpc of the galaxies 
as a function of the impact parameter.
The left panel shows the resul of model SF. The middle
panel presents the result of model SF, when 
the IGM is assumed to be homogeneously enriched 
at the level of [C/H]=-2.5 and [O/H]=-3. The right panel
demonstrates the result of model SF, when 
the QSO only UV background radiation suggested by \citet{hm01}
is adopted. The gray lines indicate median and 25th and 75th 
percentile.
\label{fig-bfmpc3}}
\end{figure*}

\begin{deluxetable*}{rrrrrrrrr}
\tablecolumns{9}
\tablewidth{0pc}
\tablecaption{Properties of the cental galaxy at $z=2.43$ \label{tab-pcg}}
\tablehead{
 \colhead{Model} & \colhead{E$_{\rm SN}$} & 
 \colhead{$M_{\rm vir}$\tablenotemark{a}} & 
 \colhead{$r_{\rm vir}$\tablenotemark{b}} &
 \colhead{$M_{\rm gas,vir}$} & 
 \colhead{$M_{\rm DM,vir}$} & 
 \colhead{$M_{\rm star,vir}$}  &
 \colhead{$M_{200}$\tablenotemark{c}} &
 \colhead{$T_{\rm vir}$\tablenotemark{d}}  \\
 \colhead{Name} & \colhead{(erg)} &
 \colhead{($M_{\sun}$)} & \colhead{(kpc)} &
 \colhead{($M_{\sun}$)} & \colhead{($M_{\sun}$)} & \colhead{($M_{\sun}$)} &
 \colhead{($M_{\sun}$)} & \colhead{(K)}
}
\startdata
NF & 0 &  $2.1\times10^{11}$ & 57  &
$1.7\times10^{10}$ & $1.7\times10^{11}$ & $2.5\times10^{10}$ &
$2.0\times10^{11}$ & $3.8\times10^5$
\\
SF & $3\times10^{51}$ &  $2.0\times10^{11}$ & 57 &
$1.5\times10^{10}$ & $1.7\times10^{11}$ & $1.6\times10^{10}$ &
$1.9\times10^{11}$ & $3.7\times10^5$
\\
ESF & $5\times10^{51}$ &  $1.7\times10^{11}$ & 56 &
$1.2\times10^{10}$ & $1.6\times10^{11}$ & $1.7\times10^{9}$ &
$1.6\times10^{11}$ & $3.2\times10^5$
\enddata
\tablenotetext{a}{Virial Mass in the definition of \citet{ks96}}
\tablenotetext{b}{Virial radius in the definition of \citet{ks96}}
\tablenotetext{c}{Mass within a radius which is the radius of a sphere
 containing a mean density of 200 times the critical density at $z=2.43$}
\tablenotetext{b}{Virial temperature in the definition of \citet{ks96}}
\end{deluxetable*}

\section{Method}
\label{sec-meth}

\subsection{Numerical Simulations}
\label{sec-nsim}

The simulations were carried out using the Galactic Chemodynamics
Code {\tt GCD+} \citep{kg03a}.
{\tt GCD+} is a three-dimensional tree $N$-body/smoothed
particle hydrodynamics (SPH) code that incorporates self-gravity,
hydrodynamics, radiative cooling, star formation, SN
feedback, and metal enrichment. {\tt GCD+} takes into account chemical
enrichment by both Type~II (SNe~II) \citep{ww95}
and Type~Ia (SNe~Ia) \citep{ibn99,ktn00} SNe and mass loss
from intermediate-mass stars \citep{vdhg97}, 
and follows the chemical enrichment history
of both the stellar and gas components of the system.
Figure \ref{fig-tnco} shows the total number of SN (both SN II and SN Ia) and 
the total amount of carbon and oxygen ejected from a star particle 
with the mass of 1 Msun as a function of its age. Initially, SNe II go off, 
and they continue until the 8 Msun star dies ($\sim$0.04 Gyr in the case
of $Z=0.02$). There is no SN until SNe Ia start to occur around 0.7 Gyr.
A star particle with $Z=10^{-4}$ does not lead to SN Ia, because
the adopted SNIa model restricts the metallicity range for progenitors
of SN Ia to $\log Z/Z_{\sun}>-1.1$ \citep[see][for details]{ktn00}.
Oxygen is produced mainly by SN II. After SN II ceases, the continuous
ejection of oxygen and carbon is mainly due to the contribution from
intermediate-mass stars. Although oxygen yield is mainly from the pre-enriched
ejecta, carbon is newly processed in intermediate-mass stars, which
explains the significant yield in the low metallicity case.

The adopted version of the code also includes
non-equilibrium chemical reactions of hydrogen and
helium species (H, H$^{+}$, He, He$^{+}$, He$^{++}$, H$_{2}$,
H$_{2}^{+}$, H$^{-}$) and their cooling processes,
following the method of \citet{aazn97,azan97,gp98}.
The details of the non-equilibrium chemical reactions
are described in the Appendix of \citet{kacg06}. 
We have made the following update from the code used in \citet{kacg06}.
We adopt a density threshold for star formation, and
permit star formation from  gas whose hydrogen number density
($n_{\rm H}=f_{\rm H} \rho_g /m_p$, where $f_{\rm H}$, $\rho_g$ and
$m_p$ are the hydrogen mass fraction, density, and proton mass
for each gas particle) is higher than 0.01 cm$^{-3}$ \citep{js04}.
It is also crucial to take into account the effect of the UV background
radiation when studying  the properties of the IGM.
We use the UV background spectrum suggested by \citet{hm01}.
The code follows non-equilibrium chemical reactions of hydrogen and helium
species subjected to the UV background.
In addition, radiative cooling and heating due to heavy elements
are taken into account 
based on the Raymond-Smith code \citep{rs77}, used in \citet{ckor95}.
The simulation starts at $z=29.7$, and initial temperature and
the fractions of hydrogen and helium species are calculated
by {\tt RECFAST} \citep{sss99,sss00}.
We turn on the UV background radiation at $z=6$ \citep{bfw01,fnl01}.

The cosmological simulation adopts
a $\Lambda$-dominated cold dark matter ($\Lambda$CDM) cosmology 
($\Omega_0$=0.24, $\Lambda_0$=0.76,
$\Omega_{\rm b}$=0.042, $h=0.73$, $\sigma_8=0.74$, and $n_s=0.95$)
consistent with the measured parameters from three-year 
{\it Wilkinson Microwave Anisotropy Probe} data
\citep{sbd06}. We use a multi-resolution technique to achieve high-resolution
in the regions of interest, including the tidal forces from
neighboring large-scale structures.
The initial conditions for the simulations are constructed
using the public software {\tt LINGER} and {\tt GRAFIC2} 
\citep{eb01}. Gas dynamics and
star formation are included only within the relevant high-resolution
region ($\sim$6~Mpc at $z$=0); the surrounding low-resolution region
($\sim$55~Mpc) contributes to the high-resolution region only through
gravity. Consequently, the initial condition consists of a total of 
1350380 dark matter
particles and 255232 gas particles.
The mass and softening lengths of individual gas (dark matter)
particles in the high-resolution region are $7.61\times10^6$
($3.59\times10^7$) M$_{\sun}$ and 1.15 (1.93) kpc, respectively.
The high-resolution region is chosen as the region within 8 times the virial
radius of a small group scale halo with the total mass of 
$M_{\rm tot}=3\times10^{12} {\rm M}_{\sun}$ and the virial 
radius of $r_{\rm vir}=380$ kpc at $z=0$. 

 We simulate the following three models with these initial conditions 
to investigate the effect of SN feedback. Model NF is a "no-SN-feedback" model:
although the model follows the chemical evolution due to SNe and 
mass loss from stars, we ignore the effect of
energy feedback by SNe. Model SF is a "strong feedback" model,
where each SN yields the thermal energy of $3\times10^{51}$ erg.
This model produces a feedback effect noticeable in a number of observables.
In the final model, ESF, for "extremely strong feedback" model,
a thermal energy release of $5\times10^{51}$ erg per SN is assumed. 
We found that this model causes too strong effects of feedback,
and produces too few stars. Therefore, the model is obviously
unrealistic. However, we retain the model for this extreme case
to help put the other models in perspective.

 We analyze the properties of the IGM for all the models at $z=2.43$. 
As mentioned above, we adopt the multi-resolution technique.
We extract a spherical volume within the radius of
$r_{\rm p}=800$ kpc (in physical scale at $z=2.43$) from a galaxy
in the high-resolution region. The central galaxy is the biggest
galaxy in the simulation volume, and the radius is chosen 
to avoid contamination from the low-resolution particles. 
In this paper, we use the coordinate system where the central
galaxy resides at (x,y,z)$=$(0,0,0).
Figure \ref{fig-pmap} shows the gas density, temperature, and
stellar density map of the central $800\times800$ proper kpc$^{2}$ region
of this volume analyzed for all the models. Figure demonstrates
that the stronger feedback affects the gas density distribution,
and suppresses star formation more dramatically.

\subsection{The Artificial QSO Spectra}
\label{sec-aspec}

 The aim of this paper is to search for signatures of galactic
winds among the absorption lines in the background QSO spectrum. 
To this end, we construct artificial QSO spectra, 
with lines of sight through the simulation volume
from various orientation and projected positions,
and compare the absorption line features among models with
different strengths of SN feedback.
For a given line of sight we identify 
the gas particles whose projected distance is smaller than
their SPH smoothing length. 
In this paper, we focus on three absorption features, HI$_{1216}$, 
CIV$_{1548}$, and OVI$_{1032}$, and we call them HI, CIV, and OVI 
hereafter. The ionization fractions for HI, CIV, and OVI
for each gas particle are derived as follows.
The HI fraction is self-consistently calculated in our simulations,
because {\tt GCD+} follows the non-equilibrium chemical reactions of the
hydrogen and helium ions.
The CIV and OVI fractions for each gas particle are analyzed
with version 6.02b of CLOUDY, described by \citet{fkv98},
assuming the condition of optical thin and ionization equilibrium.
Here, we put in the density, temperature, and 
the abundances of different elements for the gas particles
in the simulation, and run CLOUDY adopting the same UV background
radiation as used in the simulations \citep{hm01}.
Unfortunately, we realized that even our (unrealistically) strong feedback
model cannot enrich the lower density regions in the IGM as much as what is
observed \citep{cs98,sak03,ask04}.
For example, some filaments are not enriched at all even in
models ESF as will be seen in Figure \ref{fig-slmsne5}, although quantitative
comparisons with the observational data \citep[see also][]{od06}
will be pursued in a future paper. This is likely because 
the limited resolution of our simulations is unable to resolve
the formation of smaller galaxies which form at a higher redshift
and could enrich the IGM. Thus, to mimic pre-enrichment for the
low density regions we add metals at the level of [C/H]$=-3$ and
[O/H]$=-2.5$ to all gas particles.
Once the ionization fractions are obtained, the column
densities at the LOS for each species are analyzed
for each gas particle, using the two-dimensional version of the SPH kernel.
The optical depth $\tau(v)$ profiles along the LOS are calculated
by the sum of the Voigt-absorption profiles for each particle,
taking into account their temperature and LOS velocity, $v_{i,\rm LOS}$,
which is the sum of Hubble expansion and peculiar velocity. 
The final spectra are constructed, assuming an overall signal-to-noise ratio
of 50 per 0.04 $\AA$ pixel, the read out signal-to-noise ratio 500,
and FWHM$=$6.7 km s$^{-1}$.
We stress that we take into account the difference between
carbon and oxygen abundances in our chemodynamical simulations, 
when obtaining CIV and OVI fraction.

 Note that the simulation volume analyzed is only 1.6 proper Mpc
scale at $z=2.43$. The Hubble expansion at $z=2.43$ is 
236 km~s$^{-1}$~Mpc$^{-1}$ in the adopted cosmology, and 1.6 proper Mpc
corresponds to 378 km s$^{-1}$. 
In addition, since the volume is overdensity region, the expansion 
velocity is smaller than the Hubble expansion. In Section \ref{sec-mtrans},
we analyze the mean transmissivity within 1 $h^{-1}$ comoving Mpc from
the galaxies. The velocity that corresponds to 1 $h^{-1}$ comoving Mpc is
94.2 km s$^{-1}$ at $z=2.43$, which is well within the range of the volume.
However, in the real Universe, if there are absorbers outside the volume
at the LOS and their peculiar velocity is large, they can contribute
to absorptions in the velocity range we focus on 
\citep[see][for more detailed discussion of such effect]{kwdk03}. 
Therefore, we ignore the contamination from such absorptions, 
and consider the ideal
absorption systems only by the absorbers which are spatially close.

\section{Results}
\label{sec-res}

\subsection{Absorption features around a galaxy}
\label{sec-afag}

 To study how the gas outflows from galaxies 
affect the absorption line features,
we generate two sets of QSO spectra for all the models.
In this section, we analyze the spectra whose LOS are chosen
as the $5\times5$ grid points each separated by
50 kpc (in physical scale at $z=2.43$) as projected onto the plane of the sky.
The grid is centered on a galaxy.
In the next section, we generate 1000 random LOS spectra,
and compare them among the three models.
The properties of the central
galaxy chosen for the present section
for different feedback models are summarized in Table \ref{tab-pcg}.
The total mass of the central galaxy is slightly smaller than
the estimated mass of the BX galaxies 
\citep[$M_{\rm 200}\sim6.3\times10^{11}-1.6\times10^{12}$
  M$_{\sun}$ in ][]{asp05}
from the observational studies of the IGM-galaxies connection
around $z=1.9-2.6$ \citep{assp03,ass05,ssrb06}.
However, the accurate mass range for such rest-frame UV-selected
galaxies are still unknown.
In this paper, we simply assume that 
the central galaxy is a typical UV-selected star-forming galaxies,
and our simulation volume is a typical environment for such galaxies.
This assumption will be tested in our future studies.
Figure \ref{fig-sfr} shows the history of the total 
star formation rate for the star particles within 
$r=5$ proper kpc at $z=2.43$.
The extremely strong feedback in model ESF terminates star formation
in the system. We confirmed that SNe Ia continuously heat the
ISM, which keeps maintaining an outflow in model ESF. 

 We chose the X--Y plane in Figure \ref{fig-pmap}
as the projected plane of the sky, and define the Z-axis as the LOS direction.
Figures \ref{fig-slmsne0}-\ref{fig-slmsne5} demonstrate
the overdensity, temperature, metallicity, and
abundance ratio of carbon to oxygen, [C/O], in the  X--Z plane
at three different position of Y$=\pm50$ and 0 proper kpc
for models NF, SF, and ESF, respectively.
In the same planes, Figures \ref{fig-slzmsne0}-\ref{fig-slzmsne5}
give the density distributions of HI, CIV, and OVI,
and the OVI weighted temperature map. In these figures,
the velocity field of the gas component is also shown
with tangential arrows.

In model NF the central galaxy ends up surrounded by a hot gaseous halo
with a radius of about 100 kpc (Fig.\ \ref{fig-slmsne0}).
This is  due to infalling gas being shock-heated to the virial temperature 
(Table \ref{tab-pcg}). This situation appears common for high redshift
galaxies \citep{rhs97} and corresponds to the hot
accretion mode of \citep{kkwd05,db06}

Because of the inflow, the  metallicity of the hot gas is low.
On the other hand, the high density filaments are cold, and part of the gas 
keeps accreting through the filaments onto the galaxy, i.e., 
the cold accretion mode described by \citet{kkwd05} and 
\citet{db06}
(see also the velocity field in Fig. \ref{fig-slzmsne0}).
As a result, the HI density is higher along the filament,
and gets significantly higher in the collapsed region
near the central galaxy and the neighboring galaxies.
The spatial distribution of CIV and OVI also traces the filaments,
and their densities are high in the region close to the galaxies.

 Model SF produces a more extended hot gas region than
model NF, and the hot gas is now metal-enriched
(Fig. \ref{fig-slmsne3}), compared to the NF case. Here the hot gas is
dominated  by a galactic wind
induced by strong SN feedback.
The temperature map of Figure \ref{fig-slmsne3}, the arrows 
in Figure \ref{fig-slzmsne3}, and Figure \ref{fig-dZsne3}
demonstrate that the enriched gas tends to escape
toward the lower density regions. The filament
remains unaffected by the wind.
As a result,  the cold accretion
is still maintained through the filaments, which continues
funneling gas to the galaxy. 
This is similar to what was observed in 
previous numerical simulations with strong feedback effect 
\citep[e.g.,][]{tvk02,kmco06}. Consequently, the distribution of
HI density is not significantly different from model NF.
On the other hand, higher density CIV and OVI gas extends 
over a much larger region.
This is because the enriched gas is blown out from the galaxy,
and helps to raise the abundance of carbon and oxygen in the IGM. 

 Interestingly, we find that the gas whose OVI density is high  in model SF 
has a low temperature which is consistent with photoionization
equilibrium for OVI
(typical logarithmic temperatures are around Log($T$(K))$=4.2$) 
as opposed to  collisional ionization temperature
(typical Log($T$(K))$=5.5$).
Comparison between the 3rd and 4th columns in Figure \ref{fig-slzmsne3}
demonstrates that the region where the density of OVI is high
has temperatures around Log($T$(K))$<4.5$. 
This is gas that has been blown out of the galaxy, and cooled down by radiative cooling
after colliding with the ambient IGM. CIV in low density
halo and void regions is in a similar thermal state. 

 Figure \ref{fig-slmsne5} shows that the extremely strong SN feedback
in model ESF develops a much stronger galactic wind and 
a larger hot gas bubble. 
Feedback is now strong enough to affect the filaments. 
Gas accretion through the filaments is suppressed, 
so that star formation in the central galaxy ceases (Fig.\ \ref{fig-sfr}).  
However, even here the denser regions of the filaments
survive, and the density distribution of HI
is similar to the one seen in models NF and SF. 
Figure \ref{fig-slzmsne5} reveals that in model ESF
there is more collisionally ionized OVI especially 
in the region close to the galaxy ($r\leq40$ kpc)
where the OVI weighted temperature is around Log($T$(K))$=$5.5
and the OVI density peaks.

%

 Figures \ref{fig-spgridsne0}-\ref{fig-spgridsne5}  
represent the spectra whose LOS are chosen
as the $5\times5$ grid points each separated by
50 proper kpc projected on the sky
i.e., in the X--Y plane. In the figures, the middle panels 
correspond to the LOS through the center of the galaxy.
 First, we compare HI absorption lines between the three models.
At the LOS through the central galaxy, the HI lines are heavily saturated,
i.e., they produce a damped Lyman alpha line, except in model ESF.
In model ESF, not enough cold gas can survive in the galaxy
due to the extremely strong feedback.
In all the models, the HI absorption becomes weaker with
the projected distance from the galaxy. General features
of HI absorption lines are similar among the three models.
Thus, the signature of a galactic wind
seems to be difficult to see in HI absorption lines, confirming the
conclusions of previous studies
\citep{tvk02,chsww02,bfm03,kmco06}. 

However, a more detailed comparison of HI absorption line features between
models NF and SF
(Figs.\ \ref{fig-spgridsne0} and \ref{fig-spgridsne3}, respectively)
shows that the HI absorption lines tend to be stronger in model SF
than those in model NF, even close to the galaxy. This is a 
counter-intuitive result, because
it seems natural that a strong wind should  predominantly  be destroying
HI clouds, as suggested by \citet{assp03} and \citet{bw06}. 
However, as shown in Figures \ref{fig-pmap} and \ref{fig-slmsne3},
strong feedback redistributes the high density gas in the galaxy
to the surrounding region, and the filaments become broader.
This leads to the stronger HI lines in model SF, especially 
in the outer regions of the galaxy.
A similar effect is seen in model ESF (Fig.\ref{fig-spgridsne5}).
Therefore, we suggest that stronger SN feedback actually increases the
HI absorption.  
We will test this more quantitatively in the next section.

Conversely, we do see a lack of neutral hydrogen along some LOS close
to the galaxy.
However, they do not seem to have anything to do with winds. In model
NF, the LOS at 
$({\rm X},{\rm Y})=(-100,-50)$ and $(-100,-100)$
have very weak HI absorption lines, although their projected distance 
is smaller than $\sim$120 proper kpc ($\sim300$ h$^{-1}$ comoving kpc).
\citet{ass05} studied HI absorption lines around
star-forming galaxies at $2\leq z\leq3$, and found that
some LOS which are within 1 h$^{-1}$ comoving Mpc from the galaxies
show weak or absent HI absorption.
They argue that such a lack of absorption may be caused by a
galactic superwind destroying the neutral hydrogen. 
However, our model NF does {\em not} include any SN feedback.
The LOS at $({\rm X},{\rm Y})=(-100,-50)$ corresponds to the line at X$=-100$
in the lower panels of Figures \ref{fig-slmsne0} and \ref{fig-slzmsne0}.
This LOS passes through hot accretion-shocked gas which cannot accommodate HI,
and misses the more HI-rich filaments. This example
demonstrates that it is possible to have LOS 
close to galaxies which do not show any strong HI absorption,
without the need for a galactic wind.

 It is also worth mentioning that there are some LOS which
show double HI absorption lines, especially in model NF, 
e.g., the LOS at $({\rm X},{\rm Y})=(50,0)$.
We find that this is due to symmetric gas infall from the filaments.
The LOS at $({\rm X},{\rm Y})=(50,0)$ in model NF can be seen 
at X$=50$ kpc at the middle panels of Figure \ref{fig-slzmsne0}. This LOS
passes through two filaments at Z$\sim-120$ and $\sim75$ kpc,
and the velocity map shows the filament at Z$\sim-100$ ($\sim75$)
has positive (negative) LOS velocity. As a result, these two filaments
appear as double absorption components.
Such double HI absorption line features become less obvious
in the cases of stronger feedback, because outflow from
the galaxy makes the velocity field more chaotic, and fills the gap between
the components in velocity space.

 We also compare CIV and OVI lines among the models. In model NF,
there are almost no CIV or OVI lines at R$\geq50$ proper kpc.
The LOS at $({\rm X},{\rm Y})=(50,-50)$
shows strong CIV lines. However,
this is due to the next closest galaxy, as seen in Figure \ref{fig-slzmsne0}.
In contrast, the stronger feedback in models SF and ESF creates CIV and OVI
lines further away from the galaxy. We also investigated the absorption
lines with a much finer grid of LOS, i.e., smaller separations, 
and found that strong CIV or OVI lines are rare 
at projected radii  $R\geq100$ kpc even in models SF and ESF, 
unless there is another galaxy close to the LOS. 
This can also be seen in Figures \ref{fig-slzmsne3}
and \ref{fig-slzmsne5}, where dense CIV and OVI regions
extent to about 100 kpc.
OVI lines are very rare
in model NF, and only exist where the HI absorption is saturated.
On the other hand, strong feedback models produce more OVI lines,
that are sometimes stronger than CIV lines. These figures
demonstrate that the OVI lines are the most sensitive signature
of a galactic wind in absorption. We study this possibility more quantitatively
in the next section.

\subsection{The mean transmissivity of HI, CIV, and OVI}
\label{sec-mtrans}

 In this section, we analyze the artificial QSO spectra
in 1000 random LOS. Since our simulation volume is 
a spherical volume (Sec.\ \ref{sec-aspec}), we cannot
use the LOS at too large an impact parameter.
Therefore, we generate 1000 spectra for the random LOS 
at the projected radius of $R<400$ proper kpc. We also change
the angle of projection randomly for each LOS.

 Within the three dimensional radius of $r=400$ proper kpc, 
the high-resolution volume of the simulations contains
two galaxies whose virial mass is more than $10^{11}$ $M_{\sun}$.
Since the virial mass of the observed UV-selected
galaxies is not well known, as mentioned above \citep[see also][]{ess06}, 
we assume these two galaxies are such galaxies, and apply a
similar analysis to the one done for the observed UV-selected
galaxies \citep{assp03,ass05}.
 \citet{ass05} measured the mean transmissivity of all HI pixel
in their QSO spectra that lie within 1 $h^{-1}$ comoving Mpc from
the galaxies as determined from  the projected distance 
and the LOS velocity. In this paper we indicate the mean transmissivity
as $f_{\rm 1 Mpc}$. We analyze $f_{\rm 1 Mpc}$ not only for HI 
($f_{\rm 1Mpc,HI}$)
but also for CIV ($f_{\rm 1Mpc,CIV}$) and OVI ($f_{\rm 1 Mpc,OVI}$)
for all the LOS spectra for our two galaxies.
In reality, it is difficult to measure $f_{\rm 1 Mpc}$ for metal
lines. For example, OVI lines are often contaminated by 
interloper Lyman alpha lines. However, we carry out this theoretical
exercize to understand the effect of galactic winds on absorption lines
quantitatively.

 Figure \ref{fig-bfmpc} shows the mean transmissivity
of HI, CIV, and OVI as a function of the projected distance
from the galaxy, i.e., the impact parameter, $b$.
Figure \ref{fig-fhist} shows the histogram of the probability
of the decrement which is defined as $1-f_{\rm 1 Mpc}$
from Figure \ref{fig-bfmpc}. The top panels of
Figures \ref{fig-bfmpc} and \ref{fig-fhist} correspond to
the lower panel of Figures 13 and 15 of \citet{ass05}, respectively.
The left-top panel of Figure \ref{fig-fhist} shows 
results similar to those from previous numerical simulation studies with
without feedback \citep{kwdk03,tb06}.
Although the observational data (dotted histogram) of \citet{ass05}
agree as far as strong absorption is concerned,
their data show a much higher probability for 
the very weak absorption ($1-f_{\rm 1 Mpc}<0.2$).
\citet{ass05} claim that this may be because in the real universe
galactic winds turn moderate absorption into weak absorption,
but do not affect the strong absorption systems.
However, as seen in the top panels of Figures \ref{fig-bfmpc} 
and \ref{fig-fhist}, our simulations predict that
the existence of the strong galactic wind does not change
the mean flux transmissivity of the HI lines.
Interestingly, if we compare the transmissivity at the LOS
close to the galaxy ($b\leq0.4 h^{-1}$ comoving Mpc) in
Figure \ref{fig-bfmpc}, the stronger feedback
leads to slightly stronger mean absorption, i.e., smaller median 
$f_{\rm 1Mpc,HI}$, although the difference is subtle.
Therefore, again, we conclude that HI absorption lines generally
are unaffected by galactic winds.
This leaves an inconsistency between the observations and
numerical simulations. 
Unfortunately, the current number of the observational sample
is not satisfactory \citep[31 systems in][]{ass05} to reach
firm conclusions. On the theory side, the implementation of feedback
has been one of the more uncertain ingredients in current numerical simulations
\citep[e.g.][]{oefj05,ksw06,stws06}. We adopt the simplest implementation,
but different implementations may lead to different
conclusions. Clearly, further observational studies
and numerical simulations are required to address this problem.

 Figures \ref{fig-bfmpc} and \ref{fig-fhist} also show the
results for CIV and OVI lines, which appear to be 
more sensitive to the effect of SN feedback.
Model NF shows low $f_{\rm 1 Mpc, CIV}$ to be almost independently 
of the impact parameter. On the other hand, models SF and ESF
show that significantly more LOS have a lower 
$f_{\rm 1 Mpc, CIV}$ at $b<\sim0.2 h^{-1}$ comoving Mpc.
The mean transmissivity of OVI lines also shows a similar
trend. However, for OVI the absorption becomes noticeably stronger
i.e., $f_{\rm 1 Mpc, OVI}$
decreases, as the impact parameter decreases below $b<\sim0.4 h^{-1}$
comoving Mpc.
 Figure \ref{fig-fhist} reveals that model NF 
barely shows a decrement $1-f_{\rm 1 Mpc}$ higher than 0.2 for
both CIV and OVI, but models SF and ESF can produce such strong
absorption lines.
However, note that y-axis of the figure is the logarithm
of probability, and it represents only $\sim$1 \% of $f_{\rm 1 Mpc}$
that show such strong absorption in models SF and ESF.

 In the previous section, we suggested that OVI is a good tracer
for a galactic wind, and 
in model NF OVI lines are only observable where the HI lines 
are saturated. On the other hand, models SF and ESF produce
OVI lines even where the HI is not saturated.
Figure \ref{fig-fovi} plot the $f_{\rm 1 Mpc, HI}$ against
$f_{\rm 1 Mpc, OVI}$. Figure clearly shows that models SF and ESF
show the significant fraction of the spectra with
relatively weak HI ($f_{\rm 1 Mpc, HI}>0$)
and stronger OVI ($f_{\rm 1 Mpc, OVI}<0.95$).
We have calculated the probability of 
low $f_{\rm 1 Mpc, OVI}$ for the spectra with $f_{\rm 1 Mpc, HI}>0.2$.
Model NF has only 14 \% of the spectra with $f_{\rm 1 Mpc, OVI}<0.95$,
while models SF and ESF have 24 and 28 \%. 
Although the difference is small, the stronger feedback seems
to produce more such HI weak OVI strong lines.

Finally, we briefly mention how our results are sensitive to
the distribution of metals and the assumed UV background 
radiation. Since re-simulations changing the metal yields and
the UV background radiation are computationally too expensive,
we analyzed the results of model SF, assuming different
metal distributions and UV background radiations. 
Figure \ref{fig-bfmpc3} shows the results of the same analysis as 
Figure \ref{fig-bfmpc} in the cases when the the IGM is assumed 
to be homogeneously enriched at the level of [C/H]$=-2.5$ and 
[O/H]$=-3$ (model SFhZ) and when the QSO-only UV background radiation 
suggested by \citet{hm01} is adopted (model SFQ), while model SF
includes radiation from both QSO and galaxies. Model SFhZ demonstrates that 
if the heavy elements are homogeneously distributed with the assumed
metallicity, there is very little correlation between the strength
of metal lines and the impact parameter. Therefore, it is important
to follow the metal distribution in the IGM self-consistently. Model SFQ
shows that $f_{\rm 1Mpc,CIV}$ and $f_{\rm 1Mpc,OVI}$ differ little between
the QSO and galaxy radiation and QSO-only UV background radiation 
cases, except for very subtle decrease in CIV absorptions and
increase in OVI absorptions \citep[see also][]{ash05}.

\section{Conclusions}
\label{sec-conc}

 We have analyzed the QSO absorption features obtained
from cosmological numerical simulations with
different strengths of SN feedback. Our simulations
self-consistently follow the metal exchange histories
among the IGM, ISM, and stellar components. We investigate
not only  the neutral hydrogen absorption lines
but also the ionization lines for heavy elements,
keeping track of the abundance history of the elements.

 We have paid particular attention to the properties of
the IGM around high-redshift ($z=2.43$) galaxies
with $M_{\rm vir}\sim10^{11}$ $M_{\sun}$.
We found that a model without mechanical feedback creates
hot gas halos around galaxies due to shock heating,
with radii up to 100 proper kpc.
We found that such hot gas can lead to a lack of HI absorption (Fig.\
\ref{fig-spgridsne0})
even for LOS close to galaxies,
as  found by \citet{ass05}, without having to invoke a galactic wind.

In our strong feedback models, outflows induced by SN feedback
produce larger hot bubbles around 
galaxies (Figs.\ \ref{fig-slmsne3} and \ref{fig-slmsne5}).
However, such outflows tends to escape to lower density regions,
and hardly affect the dense filaments producing HI absorption systems so that
the transmissivity of HI Lyman alpha is virtually independent
of the strength of SN feedback.
If anything the absorption by neutral hydrogen slightly increases in
the presence of a wind.
We conclude that
the presence or absence of HI absorption lines is not a good indicator of 
the presence or absence of a galactic wind.

 On the other hand, we found that the metal lines, especially
OVI, are sensitive to the existence of outflows.
Without feedback, it is difficult to enrich
the IGM enough to produce strong OVI lines
further away from galaxies (Fig.\ \ref{fig-slmsne0}), 
unless there are nearby satellite galaxies intersected by the LOS by chance.
We also found that, in the no-feedback model, strong OVI lines are
almost always
associated with saturated HI lines. On the other hand,
the strong feedback model can produce strong OVI lines even where
HI lines are unsaturated,  
because strong feedback can re-distribute the enriched gas 
to relatively low density regions.
We have confirmed this by looking for the spectra whose OVI flux is less 
than 0.8 over more than 5 pixels and whose mean HI flux within 
$\pm50$ km s$^{-1}$ from the HI velocity corresponding to the OVI lines
are higher than 0.2 from 1000 spectra with random LOS. 
The no-feedback model has no such spectra, while strong feedback 
(model SF) has 12 of such spectra.
We point out that Figure 9 of \citet{ssr04}
shows an OVI line where the HI is not saturated.
Our results suggest that this is likely to be a region
where the effect of a galactic wind is significant.
Analyzing the transmissivity of OVI lines we found that
strong feedback creates more LOS with lower transmissivity, i.e.
stronger OVI absorption, near the star-forming galaxies. 
The statistical analysis of transmissivity also shows that there are more LOS
where stronger OVI is associated with weaker HI, 
in the presence of galactic winds.
We expect that the pixel-optical depth analysis of OVI 
against HI \citep{srsk00} would be sensitive to the presence of a
galactic wind, and we will test this idea in a future paper.
In conclusion, OVI appears a theoretically good tracer of galactic winds
that merits further attention.

\acknowledgments

DK thanks the financial support of the JSPS, through 
Postdoctoral Fellowship for research abroad. 
We acknowledge 
the Center for Computational Astrophysics of the National Astronomical
Observatory, Japan (project ID: imn33a), the
Institute of Space and Astronautical Science 
of Japan Aerospace Exploration Agency, and
the Australian and Victorian Partnerships for Advanced
Computing, where the numerical computations for this paper were
performed. MR is grateful to the NSF for support under grant AST-05-06845.


\end{document}